\documentstyle[12pt]{article}

\newcommand{\beq}{\begin{equation}}
\newcommand{\eeq}{\end{equation}}
\begin{document}
\title
{The Isospin Splittings of Heavy-Light Quark System}

\author{Leonard S. Kisslinger and Zhenping Li\\
      Department of Physics, Carnegie Mellon University\\
      Pittsburgh, PA 15213}

\maketitle
\indent
\begin{abstract}
The mass splittings of the pseudoscalar and vector D and B light-heavy
quark systems have been calculated using the method of QCD sum rules.
Electromagnetic, quark mass, and nonperturbative QCD effects are all
included.  The results are in good agreement with experiment.  A
measurement of isospin splitting for the vector B mesons would
 give valuable information about quark mass splittings.
\vspace{3mm} \\

\end{abstract}
\newpage
The origin of mass differences in isospin multiplets
has long been of great  interest in nuclear and particle
physics as a source of information about symmetry violations.
Hadronic isospin violations are particularly important
in that they arise from nonperturbative Quantum
Chromodynamics (QCD) as well as quark mass differences
(see Ref.\cite{gl} for a review of the early work in
this area), and of course electromagnetic effects.
Among the first applications of the method of QCD
sum rules was the study of isospin violations in
the $\rho-\omega$ system\cite{svz}, in which it
was recognized that the isospin splitting of the
light-quark condensates can produce effects as
large as the current-quark mass splittings and
electromagnetic effects.  Recently the QCD sum rule
 method has been used to study the neutron-proton
mass difference\cite{hhp,yhhk}, the octet baryon
mass splittings\cite{adi} and the mass differences in
the charmed meson systems (the D and $\rm{D}^*$ scalar
and vector mesons)\cite{ei}.

In the present work we study the mass splittings of
the D,$\rm{D}^*$, B and $\rm{B}^*$ systems using the
QCD sum rule method.  A main feature of the study
is a consistent treatment of electromagnetic corrections
within the framework of the method.  Rather than make
use of estimates of electromagnetic effects from quark models
as used in previous work, or a phenomenological parametrization of
electromagnetic effects as in Ref\cite{yhhk}, we explicitly
calculate the two-loop electromagnetic
corrections as well as the perturbative and nonperturbative QCD processes
including quark masses up to dimension eight.  This is part of a
study of light-heavy mesons\cite{kl} in which the accuracy of heavy
quark effective theories is tested.

The isospin splittings are particularly
interesting in that the electromagnetic effects and quark mass splittings
enter with different relative signs in the charmed vs the bottom systems,
magnifying the role of the isospin violation of the quark condensates.
On the other hand, nonperturbative QCD mechanisms are expected to be less
important for heavy quark systems, so that an accurate treatment of
electromagnetic processes allows one to study light quark mass differences.
That will be our main conclusion.

One can write the correlators in the QCD sum rule as
\beq\label{1}
\Pi_t(p^2)=\Pi_0(p^2)+m_q\Pi_m(p^2)+\frac {\alpha_{e}}{4\pi}
 \Pi_{em}(p^2).
\eeq
For $\Pi_0(p^2)$, the leading term for the light heavy quark system,
the isospin violation comes from the isospin splitting of the
quark condensates in the non-perturbative power correction, which
has been given in previous publications\cite{kl,AE83}. The second
term in Eq. \ref{1} is the light quark mass, $m_q$, expansion of the
total correlator;  and the third term corresponds to the electromagnetic
corrections, which have been not treated consistently in the framework of the
QCD sum rule method in the literature.  The $m_q$ expansion
is extended to dimension 5 in our calculation, and furthermore, we also
include the one-loop corrections to the quark condensate, which we believe
is significant.  Thus, $\Pi_m(p^2)$ in Eq. \ref{1} can be written as
\begin{equation}\label{2}
\Pi_m(q^2)=C_I I + C_3 \langle\bar q q\rangle
+C_5 \langle  \bar q(\sigma \cdot G) q\rangle
\end{equation}
where the coefficient $C_I$  represents the perturbative contributions,
whose Feynman diagrams are shown in Fig. 1.  Based on the results in Ref.
\cite{gb}, we find the $m_q$ expansion to order $\alpha_s$ is
\begin{equation}\label{3}
Im C_I^{p.s}=\frac {3M_Q}{4\pi} (1-x)\left [1+\frac {4\alpha_s}
{3\pi}\left (f(x)+\frac 34x+\frac 32 x
\ln \left(\frac {x}{1-x}\right )\right )\right ]
\end{equation}
for the pseudoscalar and
\begin{equation}\label{4}
Im C_m^{v}=Im C_m^{p.s}-\frac {\alpha_sM_Q}{\pi^2} (1-x)
\end{equation}
for the vector correlator, where $x=\frac {M^2_Q}{p^2}$,
\begin{equation}\label{5}
f(x)=\frac 94+2l(x)+\ln(x)\ln(1-x)+
\left (\frac 52-x-\frac 1{1-x}\right )\ln(x)-\left (\frac 52-x\right )\ln(1-x)
\end{equation}
and $l(x)=-\int_0^x \ln(1-y)\frac {dy}{y}$ is the Spencer function.
Particular attention has been paid to the $m_q$ expansion of the
quark condensate since we find that it plays an important in the
heavy-light quark systems.  In particular, we have considered the
one-loop corrections to the quark condensate whose Feynman diagrams
are shown in Fig. 2-a, b, c, and d,  the result to the order $\alpha_s$ is
\begin{eqnarray}\label{6}
C^{p.s}_3=&-&\frac 12\frac 1{p^2(1-x)}\bigg \{\frac x{1-x}-1 \\ \nonumber
&+&\frac {4\alpha_s}{3\pi}\left [\frac {3x}{4(1-x)}\left (7+2\ln\left (
\frac {x}{x-1}\right )\right )
-x\left (2-(3+2x)\ln\left(
\frac x{x-1}\right )\right )\right ]\bigg \}
\end{eqnarray}
for the pseudoscalar and
\begin{eqnarray}\label{7}
C^{v}_3=&-&\frac 12\frac 1{p^2(1-x)}\bigg \{\frac x{1-x} \\ \nonumber
&+&\frac {4\alpha_s}{3\pi}\left [\frac {x}{4(1-x)}\left (1-6\ln\left (
\frac {x}{x-1}\right )\right )
+2x\left (1-x\ln\left(
\frac x{x-1}\right )\right )+\frac 12\right ]\bigg \}
\end{eqnarray}
for the vector currents.  The details of the one loop calculations
will be given later\cite{kl}.  However,
we find that the $m_q$ expansions of the one loop corrections to the
quark condensate are finite after the mass renormalization.
The $m_q$ expansion for the quark gluon condensate corresponding
to Fig. 2-e and f is
\begin{equation}\label{8}
C_5^{p.s}=\frac {x}{p^4(1-x)^3}\left(\frac {3x}{2(1-x)}-1\right )
\end{equation}
and
\begin{equation}\label{9}
C_5^{v}=\frac {1}{p^4(1-x)^2}\left ( \frac {3x^2}{2(1-x)^2}+\frac
{5x}{6(1-x)}-\frac 5{12}\right )
\end{equation}
for the pseudoscalar and vector current, respectively.

For the electromagnetic corrections, we write $\Pi_{e.m}(q^2)$
as
\begin{equation}\label{10}
\Pi_{e.m}(q^2)=D_I I + D_3 \langle\bar q q\rangle +\dots
\end{equation}
The coefficient $D_I$ in Eq. \ref{10}
is a two-loop perturbative contribution
which replace the gluon lines in Fig. 1 by the corresponding
photon lines.  Therefore, the imaginary part
of the coefficient $D_I$ can be obtained
from the two-loop integral\cite{gb} in QCD for the charge
neutral current, where the charges for the heavy quark $e_Q$ and the light
quark $e_q$ are equal with opposite sign:
\begin{equation}\label{11}
ImD_I^{p.s}=\frac {3e_q^2M_Q^2}{2\pi}(1-x)^2f(x)
\end{equation}
and
\begin{eqnarray}\label{12}
Im D_I^{v}=\frac {e_q^2p^2}{2\pi}(1-x)^2\bigg [ (2+x)(1+f(x))-(3+x)(1-x)
\ln\left (\frac x{1-x}\right )\\ \nonumber
-\frac {2x}{(1-x)^2}\ln(x)-5-2x-\frac {2x}
{1-x}\bigg ]
\end{eqnarray}
for the pseudoscalar and the vector currents,
where $f(x)$ is given by Eq. \ref{5}.
The calculation of $D_I$ for the charge currents is more complicated.
Since the charges for the heavy and light quarks are not the same,
the Ward identity can not be used here; thus, one should do
the wavefunction and mass renormalizations at the same time.
We can write the imaginary part of the correlator for
the charged currents as
\begin{equation}\label{13}
Im D_I^+=-e_Qe_q D^++(e_Q^2+e_Qe_q)D_Q+(e_q^2+e_Qe_q)D_q
\end{equation}
where $D^+$ can be obtained from Eqs. \ref{10} and \ref{11};
\begin{equation}\label{14}
D^+=D^{p.s, v}_I/e_q^2.
\end{equation}
$D_{Q,q}$ is the integral for
the self energy diagrams for the heavy and the
light quarks.  The corresponding imaginary parts are
\begin{eqnarray}\label{15}
Im D_Q^{p.s}=\frac {3p^2}{2\pi}\bigg [\frac 13(1-x)^4\left (1+
\frac 15(1-x)\right ) -\frac {17x}4\left (\ln(x)+1-x\right )\\ \nonumber
-x(1-x)^2\left (\frac {15}8-\frac 14 \ln \left (
\frac {M_Q^2}{\mu^2}\right )+\frac 5{4(1-x)}\ln(x)-\ln\left (
\frac x{1-x}\right )\right )\bigg ],
\end{eqnarray}
\begin{eqnarray}\label{16}
Im D_Q^{v}=\frac {3p^2}{2\pi}\bigg [\frac 19(1-x)^4\left (1+
\frac 15(1-x)\right ) -\frac {5x}{12}\left ((1-x)^2+2(\ln(x)+1-x)
\right )\\ \nonumber
-\frac 1{3}(1-x)^2(2+x)\left (
\frac {5}8-\frac 14 \ln \left (
\frac {M_Q^2}{\mu^2}\right )
+\frac 5{4(1-x)}\ln(x)-\ln\left (
\frac x{1-x}\right )\right )\\ \nonumber
-\frac{7x}{12}(2+x)(1-x+\ln(x))\bigg ],
\end{eqnarray}
\begin{eqnarray}\label{17}
Im D_q^{p.s}=\frac {3p^2}{2\pi}\bigg [ \frac 16
(1-x)^4\left (1+\frac 15(1-x)\right )+\frac {5x}4
\left (\frac 12(1-x)^2+x(1-x+
\ln(x))\right ) \nonumber \\
-\frac x4(1-x)^2\left (2-\ln \left (
\frac {M_Q^2}{\mu^2}\right )+2\ln \left (\frac x{1-x}\right )
-\frac 1{1-x}\ln(x)\right )\bigg ]
\end{eqnarray}
and
\begin{eqnarray}\label{18}
Im D_q^{v}=\frac {3p^2}{2\pi}\bigg [\frac {(1-x)^4}{18}\left (
1+\frac {(1-x)}{5}\right )
+\frac {(5x+2)}{12}\left ( \frac {(1-x)^2}{2}
+x(1-x+\ln(x))\right ) \nonumber \\
-\frac {(2+x)}{12}(1-x)^2\left ( 2-\ln \left (
\frac {M_Q^2}{\mu^2}\right )+2\ln \left (\frac x{1-x}\right )-\frac
1{1-x}\ln (x)\right ) \bigg ]
\end{eqnarray}
for the pseudoscalar and vector currents, respectively.

The $D_3$ in Eq. \ref{10} represents the one-photon-loop corrections
to the quark condensate, whose Feynman diagrams are shown in Fig. 2.
For the charge-neutral systems, with the appropriate change in coupling
constants, the results are
the same as the one-gluon-loop corrections to the
quark condensate which have been calculated explicitly\cite{kl}.
We find
\begin{equation}\label{19}
D_3^{p.s}=e_q^2\frac{x}{M_Q(1-x)}
 \left [2+3\ln \left (\frac
{M_Q^2}{\mu^2}\right )
-6(1-x)\left (1+x\ln\left (\frac {x-1}{x}\right )\right )\right ]
\end{equation}
and
\begin{equation}\label{20}
D_3^{v}=e_q^2\frac {x}{M_Q(1-x)}\left [
3\ln\left (\frac {M_Q^2}{\mu^2}\right )-2+2(1-x)\left (1+x\ln \left (
\frac {x-1}{x}\right )\right )\right ]
\end{equation}
for the pseudoscalar and vector respectively.  We set
$\mu=M_Q$ in our numerical evaluations.
There are different expressions for the charged $D_3$'s.
We find
\begin{equation}\label{21}
D_3^{p.s}=\frac {2}{9}\frac {x}{M_Q(1-x)}\left [ 1+
6\ln\left (\frac {M_Q^2}{\mu^2}\right )+6(1-x)
\left (1+x\ln\left (\frac {x-1}{x}\right )\right )\right ]
\end{equation}
and
\begin{equation}\label{22}
D_3^{v}=\frac {2}{9}\frac {x}{M_Q(1-x)}\left [
5+6\ln\left (\frac {M_Q^2}{\mu^2}\right )
-2(1-x)\left (1+x\ln\left (\frac {x-1}{x}\right )\right )\right ]
\end{equation}
for the charged pseudoscalar and vector currents of $D$ mesons,
and
\begin{eqnarray}\label{23}
D_3^{p.s}=\frac {2}{9}\frac {x}{M_Q(1-x)}\bigg
[4+21\ln \left (\frac {M^2_Q}{\mu^2}\right )+9(1+x)\ln\left(\frac {x-1}
{x}\right )\nonumber \\
+6(1-x)\left (1+x\ln\left (\frac {x-1}{x}\right )\right )\bigg ]
\end{eqnarray}
and
\begin{eqnarray}\label{24}
D_3^{v}=\frac {2}{9}\frac {x}{M_Q(1-x)}\bigg
[8+21\ln \left (\frac {M^2_Q}{\mu^2}\right )+9(1+x)\ln\left(\frac {x-1}
{x}\right )\nonumber \\
-2(1-x)\left (1+x\ln\left (\frac {x-1}{x}\right )\right )\bigg ]
\end{eqnarray}
for the charged pseudoscalar and vector currents of the B mesons.
There are also corresponding one-photon-loop corrections
to the quark-gluon condensates and higher dimension power corrections.
Whether they are important numerically are remained to be studied.
As we indicate later, the one photon loop corrections to the
quark-gluon condensate might be important.

Defining the quantity $\omega^2=M_Q^2-p^2$ for the
Borel transformation, where $\omega^2$ measures the
off-shell effects of the heavy quarks, we find
\begin{equation}\label{25}
f^2_p\frac {M^4_p}{M_Q^3}e^{-\frac {M^2_p
-M_Q^2}{\omega_B^2}}=\Pi_t^p(\omega_B^2)
\end{equation}
and
\begin{equation}\label{26}
f_v^2\frac {M_v^2}{M_Q}e^{-\frac {M_v^2-M_Q^2}{\omega_B^2}}
=\Pi_t^v(\omega^2_B)
\end{equation}
for the pseudoscalar and vector current. The correlator
$\Pi_t(\omega_B^2)$
can be divided into the perturbative and nonperturbative parts:
\begin{equation}\label{27}
\Pi_t(\omega_B^2)=\int_0^{\omega_0^2} Im C_I^t(\omega^2)e^{-\frac
{\omega^2}{\omega_B^2}}d\omega^2+\Pi_{np}(\omega_B^2),
\end{equation}
where the perturbative term is written in dispersion integral form,
and the nonperturbative correlator $\Pi_{np}(\omega_B^2)$ comes
from the standard Borel transformations of the terms associated
with the condensates.
The masses $M_p$ and
$M_v$ in Eqs. \ref{25} and \ref{26} can be obtained
from
\begin{equation}\label{28}
M^2=M_Q^2+\omega_B^4\frac {d\ln \left (\Pi_t(\omega_B^2)\right )
}{d\omega_B^2}.
\end{equation}
 The mass differences between different isospin states are obtained
by adjusting the parameters $\omega^2_0$ in Eq. \ref{27} to insure that
the mass difference is independent of Borel parameter $\omega_B^2$.
The range of $\omega_B^2$ is chosen in such way so that the
the continuum and the nonperturbative contributions
are minimum.

There are two parameters that are important for the isospin
mass splittings: the up and down quark mass difference, $m_d-m_u$,
and the isospin splitting of the quark condensate, which is defined
by the parameter $\gamma$;
\begin{equation}\label{29}
\gamma=\frac {\langle \bar {d}d\rangle}{\langle \bar {u}u\rangle}-1.
\end{equation}
The quark mass difference has been estimated via broken chiral
symmetry\cite{gl},
while the parameter $\gamma$ in various estimates
ranges from $-0.0079$\cite{yhhk} to $-0.002$\cite{adi,ei}.
Physically, the sensitivity of the isospin mass splittings
to the parameter $\gamma$ measures the strength of the
nonperturbative effects on the
light-heavy quark system.   To examine the sensitivity of the mass
splittings on the parameters $\gamma$,
we calculated the mass splittings for different
$\gamma$'s in our framework, with the quark mass difference
$m_d-m_u$ fixed at $3.8$ MeV.  Our result for different
$\gamma$ and $m_d-m_u$ fits approximately
to the empirical formula [all mass differences are given in MeV]
\begin{equation}\label{34}
M_{D^\pm}-M_{D^0}=0.1-50\gamma+1.1 (m_d-m_u),
\end{equation}
\begin{equation}\label{35}
M_{D^{*\pm}}-M_{D^{*0}}=-0.9-10\gamma+1.0(m_d-m_u),
\end{equation}
for D and $D^*$ mesons and
\begin{equation}\label{36}
M_{B^{\pm}}-M_{B^{0}}=1.45-1.16 (m_d-m_u)
\end{equation}
\begin{equation}\label{37}
M_{B^{*\pm}}-M_{B^{*0}}=1.36-1.28 (m_d-m_u)
\end{equation}
for B and $B^*$ mesons.
Comparing to the corresponding formula for the
neutron-proton mass difference\cite{yhhk}, neglecting electromagnetic
corrections,
\begin{equation}\label{42}
M_n-M_p=-416\gamma+0.197(m_d-m_u),
\end{equation}
we find that the isospin mass splittings are
much less sensitive to the $\gamma$ for the $D$ mesons,
and the mass splittings for the B systems is almost independent
of $\gamma$.   This result differs from the conclusion\cite{ei}
of Eletsky and Ioffe who found a stronger dependence
of isospin mass splittings for the D mesons on the
parameter $\gamma$.  The difference is due to
their neglect of higher dimension terms, especially the
quark gluon condensate in $\Pi_0(p^2)$ and $\Pi_m(p^2)$ of Eq. 1,
which is very important in our calculations.   The weak dependence of
the isospin splittings on the isospin splitting of the quark condensate
is consistent with conclusions of heavy quark symmetry that
the nonperturbative effects become less important for
heavier quark systems.   Therefore, the heavy-light quark systems
provide an excellent testing ground for the up and down
quark mass difference if electromagnetic corrections are calculated
accurately.

For $\gamma=-0.0079$ and $m_d-m_u=3.8$ MeV, we have
\begin{equation}\label{38}
M^{\pm}-M^0=\left \{ \begin{array}{r@{\quad\quad}l}
4.68 & \mbox{ for D} \\ 2.94 &\mbox{ for $D^*$}
\end{array}\right .
\end{equation}
and
\begin{equation}\label{39}
M^{\pm}-M^0=\left \{ \begin{array}{r@{\quad\quad}l}
-3.0 & \mbox{ for B} \\ -3.5 &\mbox{ for $B^*$}
\end{array}\right . .
\end{equation}
This is in excellent agreement with data
($M_{D^{\pm}}-M_{D^0}=4.7 \pm 0.7$ MeV and
$M_{D^{*\pm}}-M_{D^{*0}}=2.9 \pm 1.3$ MeV\cite{pdg})
for D mesons, while the magnitude of our result
for $B$ mesons is quite large comparing to the data
$-0.1 \pm 0.8$ MeV\cite{pdg}.
One of the important features of our calculation
is the differential vector-pseudoscalar isospin mass splittings
$\Delta (M)$;
\begin{equation}\label{40}
\Delta (M)= (M^{*\pm}-M^{*0})-(M^{\pm}-M^0),
\end{equation}
which provides an very important probe of the hyperfine
splittings of the heavy-light quark potential.
Our result indicates that $\Delta(M)$ is less dependent
on the quark mass difference $m_d-m_u$, and from
Eqs. \ref{38} and \ref{39}, we have
\begin{equation}\label{41}
\Delta(M)=\left \{ \begin{array}{r@{\quad\quad}l}
-1.7& \mbox{ for D mesons} \\ -0.5 &\mbox{ for B mesons}
\end{array}\right .
\end{equation}
for $m_d-m_u=3.8$ MeV.
This is in good agreement with recent studies by Cutkosky
and Geiger\cite{cg}.  In particular, the $\Delta (M)$
for B mesons is negative for positive $m_d-m_u$, which
is opposite to the
quark model predictions and several other investigations\cite{gi}.
  A measurement of the isospin mass splittings for the
vector B mesons will provide a crucial test in this regard.
  A study in chiral perturbation
theory finds\cite{jenkins} that the contributions
from quark mass difference only gives 0.3 MeV for
D mesons, and the rest might comes from
the electromagnetic corrections.  Our calculation
is consistent with this conclusion, and furthermore
we also find that there is only a small contribution from
the difference of the quark condensates to $\Delta(M)$.

We find the electromagnetic corrections to the
isospin splittings to D and $D^*$ mesons are quite small.
They contribute less than 1 MeV.  As a result, our splittings
in the charmed systems are too small in comparison with
experiment for parameters giving good fits to the B splittings.
One source of this discrepancy might be that
the one photon loop corrections to the quark gluon
condensate are needed, since we find that the  quark gluon
condensate is as important as the quark condensate
in evaluating the masses of the heavy ligth systems. This
might both help to explain the electromagnetic corrections
are small for the D mesons and improve our fit to the B
mass splitting with $m_d-m_u$ about 3.0 MeV. This is a somewhat
subtle calculation, which we do not attempt here.

In summary, we show how to include the electromagnetic
corrections in the QCD sum rule analysis for the
light-heavy quark systems.  This approach can also be extended
to the study of the neutron-proton mass difference as well
as other isospin mass splittings.
Our results show that the mass splittings are not
sensitive to the isospin splitting of the quark condensate, and thus
heavy-light quark systems are
an ideal place to study the  up and down quark mass differences.
We expect the method will give a very good qualitative even quantitative
descriptions of the isospin-splittings for the
 light-heavy quark systems.  Our result for $\Delta(M)$
is in good agreement with recent phenomenological studies
by Cutkosky and Geiger.    Further
investigation of the electromagnetic effects of higher
dimensions terms is needed, and this is in progress. Our
investigation shows that the isospin mass splittings of the
light heavy quark systems provide an very
important opportunity to study the light quark masses and
the hyperfine interactions of the quark-quark potential.

This work is supported in part by National Science Foundation grant
PHY-9023586.

\vspace{5mm}
\noindent {\Large\bf Figure Captions}

\vspace{5mm}
\begin{itemize}
\begin{enumerate}
\item Processes for two loops correction.
\item Contribution from nonperturbative processes.
\end{enumerate}
\end{itemize}
\end{document}